\documentclass[twocolumn]{webofc}

\usepackage[varg]{txfonts}   
\usepackage{hyperref}
\usepackage{url}
\hypersetup{colorlinks=true,citecolor=blue,urlcolor=blue,linkcolor=blue}
\begin{document}
\title{Anomaly Detection Based on Machine Learning for the CMS Electromagnetic Calorimeter Online Data Quality Monitoring
}

\author{
  \firstname{Abhirami} \lastname{Harilal}\inst{1}\fnsep\thanks
  {\email{aharilal@andrew.cmu.edu}} \and
  \firstname{Kyungmin} \lastname{Park}\inst{1}\fnsep\thanks
  {\email{kyungmip@andrew.cmu.edu}} \and
  \firstname{Manfred} \lastname{Paulini}\inst{1}\fnsep\thanks
  {\email{paulini@andrew.cmu.edu}}
  \lastname{(On behalf of the CMS Collaboration)}
}

\institute{Carnegie Mellon University, Pittsburgh, Pennsylvania, USA} 

\abstract{
  A real-time autoencoder-based anomaly detection system using
  semi-supervised machine learning has been developed for the online
  Data Quality Monitoring system of the electromagnetic calorimeter of
  the CMS detector at the CERN LHC. A novel method is introduced which
  maximizes the anomaly detection performance by exploiting the
  time-dependent evolution of anomalies as well as spatial variations in
  the detector response.  The autoencoder-based system is able to
  efficiently detect anomalies, while maintaining a very low false
  discovery rate.  The performance of the system is validated with
  anomalies found in 2018 and 2022 LHC collision data. Additionally, the
  first results from deploying the autoencoder-based system in the CMS
  online Data Quality Monitoring workflow during the beginning of Run\,3
  of the LHC are presented, showing its ability to detect issues missed
  by the existing system.  }

\maketitle
\section{Introduction}
\label{intro}
The central feature of the CMS experiment recording proton-proton
collision data produced by the CERN LHC is a superconducting
solenoid of 6\,m internal diameter, providing a magnetic field of
3.8\,T. Within the solenoid volume are a silicon pixel and strip
tracker, a lead tungstate crystal electromagnetic calorimeter (ECAL),
and a brass and scintillator hadron calorimeter, each composed of a
barrel and two endcap sections.
Muons are measured in gas-ionization detectors embedded in the steel
flux-return yoke outside the solenoid. A more detailed description of
the CMS detector, together with a definition of the coordinate system
used and the relevant kinematic variables, can be found in
Ref.~\cite{CMS:2008xjf}.

\begin{figure}[tb]
\centering{
\includegraphics[width=0.45\textwidth]{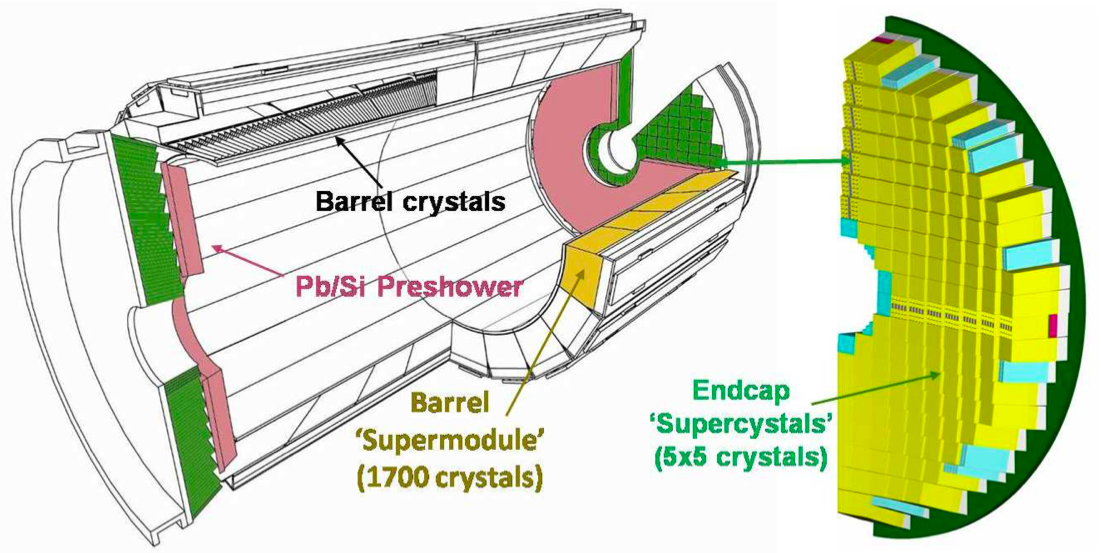}
}
\caption{Schematic view of the ECAL showing the cylindrical barrel closed
  by the two endcap regions with one half endcap displayed.}
\label{fig:ECAL}
\end{figure}

The CMS electromagnetic calorimeter provides homogeneous coverage in
pseudorapidity $|{\eta}|<1.48 $ in a barrel region (EB) and $1.48 <
|{\eta}| < 3.0$ in two endcap regions (EE$+$ and EE$-$) as shown in
Fig.~\ref{fig:ECAL}. Preshower detectors consisting of two planes of
silicon sensors interleaved with three radiation lengths of lead are
located in front of each endcap detector. The ECAL consists of 75\,848
lead tungstate (PbWO$_4$) crystals. The barrel granularity is 360-fold
in $\phi$ and (2×85)-fold in $\eta$ provided by a total of 61\,200
crystals, with each crystal having a dimension of 0.0174$\times$0.0174
in $\Delta \eta\times\Delta \phi$ space, while each endcap is divided
into two halves, with each comprising 3662 crystals. 

The CMS data quality monitoring (DQM) system~\cite{DQM2019_Azzolini} is
a crucial operational tool to record high-quality physics data.
Presently, the DQM consists of a software system that produces a
set of histograms that are based on a preliminary analysis of a subset
of data collected by the CMS detector. Conventional cut-based thresholds
are used to define quality flags on these histograms which are monitored
continuously by a DQM shifter in the CMS control room who reports on any
apparent irregularities observed.  While this system has proven to be
dependable, the changing running conditions and increasing LHC collision
rates, together with aging electronics, bring forth failure modes that
are newer and harder to predict.

\begin{figure}[tb]
\centering{
\includegraphics[width=0.30\textwidth]{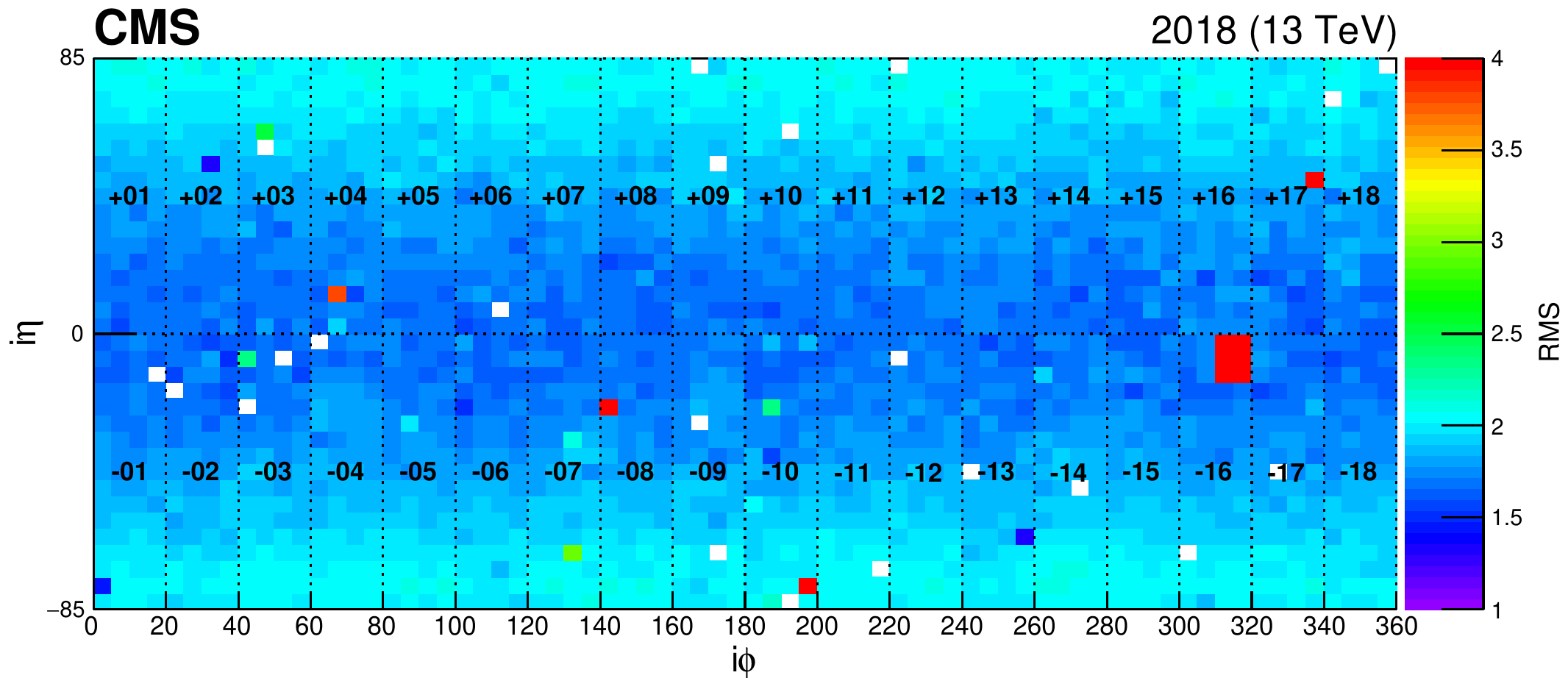} 
\includegraphics[width=0.16\textwidth]{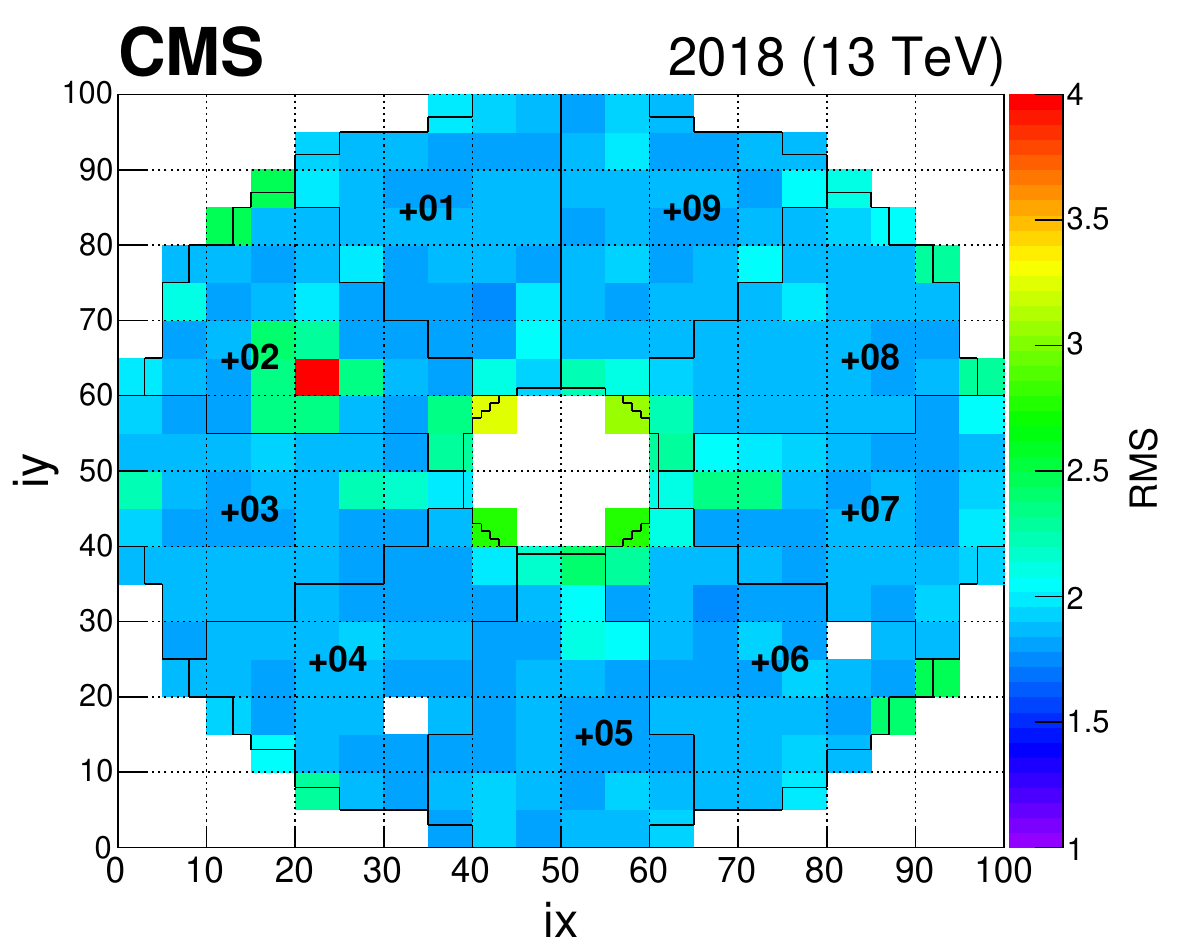} 
\includegraphics[width=0.30\textwidth]{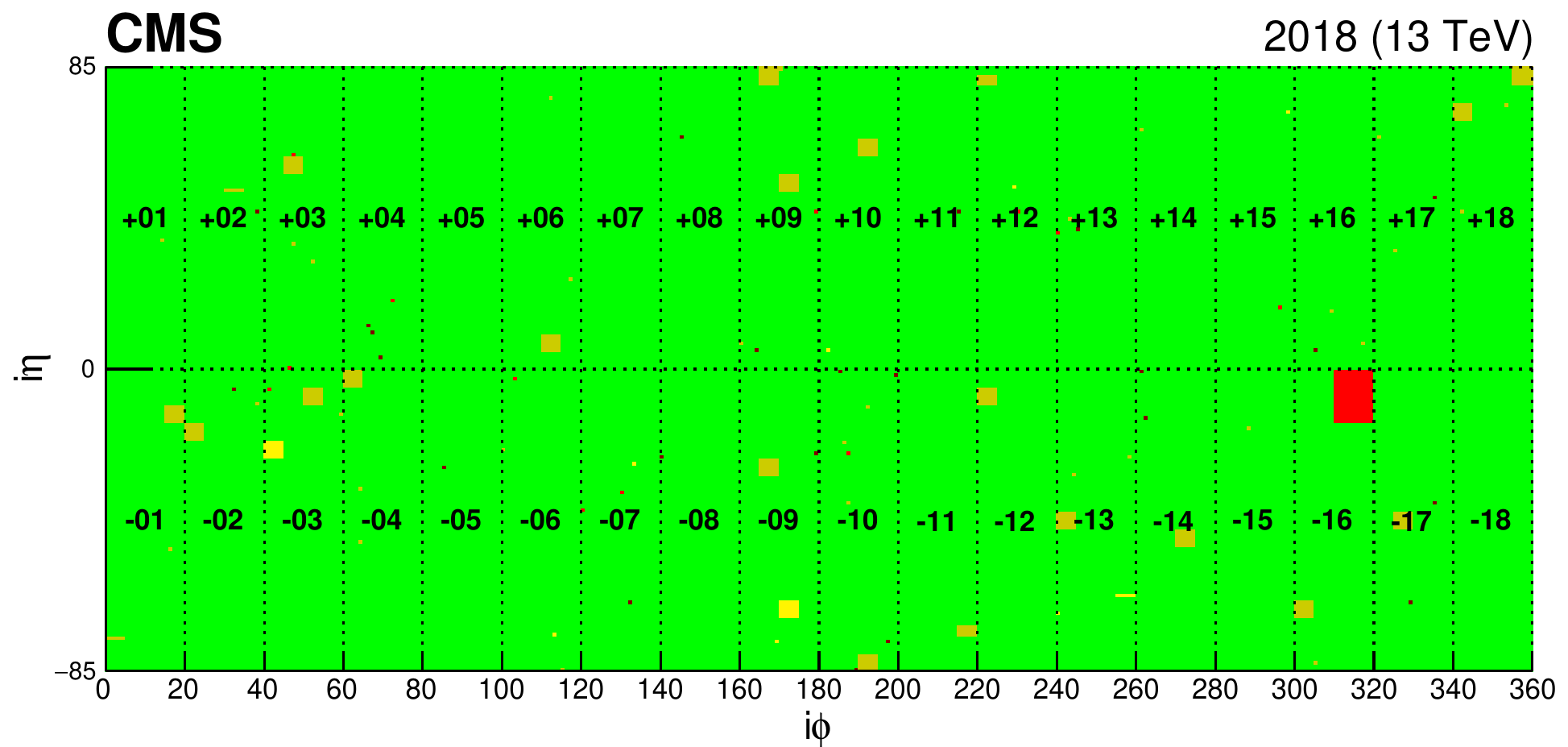} 
\includegraphics[width=0.16\textwidth]{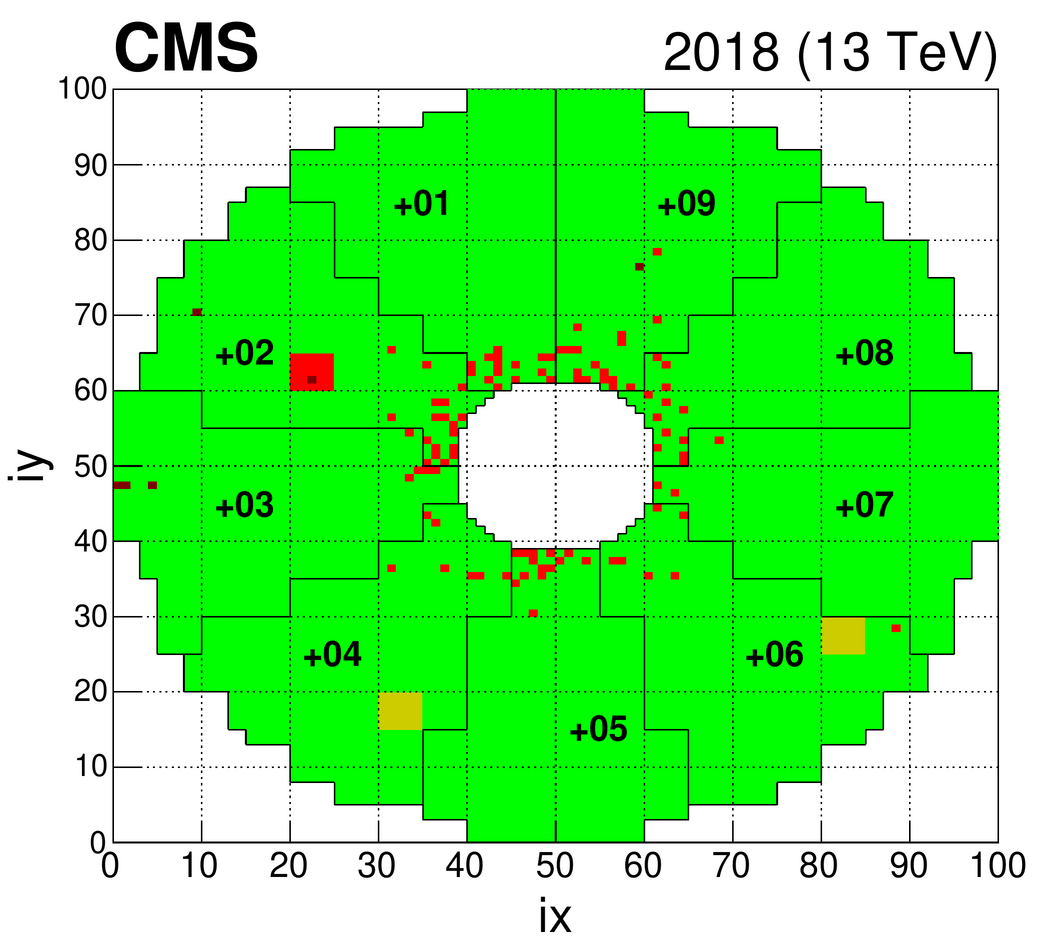}
}
\caption{Example ECAL DQM histograms showing the distribution
  of RMS of the pedestal values in the barrel and EE$+$ (top).
  The diagrams at the bottom show the
  corresponding quality map for the two regions, drawn at a
  channel-level granularity, after a set of cuts is applied on the noise
  values shown at the top.} 
\label{fig:task}
\end{figure}

There are two kinds of histograms present in the ECAL DQM:
``Occupancy-style'' histograms shown at the top of 
Fig.~\ref{fig:task} filled with critical
quantities from the real-time detector data and ``Quality-style''
histograms displayed at the bottom of 
Fig.~\ref{fig:task}. Quality-style
histograms are obtained by applying predefined thresholds and
requirements to the Occupancy-style histograms, where the thresholds are
derived from typical detector response.
The quality histograms are drawn in easily identifiable colored maps,
and the color code scheme used is, e.g., green for ``good'' and red
for ``bad'', or brown for ``known problems''.

\section{Machine Learning Based Anomaly Detection Strategy}
\label{ML-sec2}

Machine learning (ML) techniques are nowadays widely used in high-energy
physics~\cite{ML_whitepaper} and provide an excellent tool for anomaly
detection in particle physics searches~\cite{Nachman-anomalydetection}.
In this paper, an semi-supervised method of anomaly detection for the
ECAL online DQM is presented, exploiting an autoencoder
(AE)~\cite{AE} on ECAL data to supplement the DQM system. 
The network is trained exclusively on
a certified good physics dataset, so that it learns the patterns of good
data and is able to detect anything that differs from the nominal
patterns it has learned.  The network is able to detect anomalies
without the need to explicitly see the anomalous data during training.
The dataset used for training and validation of the AE
network is taken from CMS runs collected in 2018 during LHC Run\,2.
Each input image for the AE is the occupancy map from a single time
interval called ``lumi-section'' (LS) of an approximate time duration of
23 seconds.

Using an AE network based on a computer vision technique, the ML system
is built with a convolutional neural network (CNN)
architecture~\cite{lecun:98} exploiting ECAL data processed as 2D
images. The encoder part of the AE takes the input data and compresses
it into a lower dimensional representation, called the latent space,
which contains a meaningful internal representation of the input
data. The decoder part then decompresses the encoded data back to the
original image of the same dimensions, or reconstructs the image.  To
measure how well the output matches the input, a reconstruction
loss~($\mathcal{L}$) is computed using Mean Squared Error between the
input ($x$) and the AE-reconstructed output ($x'$) defined as
$\mathcal{L}(x,x') = ||(x-x')||^{2}$.

\begin{figure}[hb]
\centering{
\includegraphics[width=0.45\textwidth]{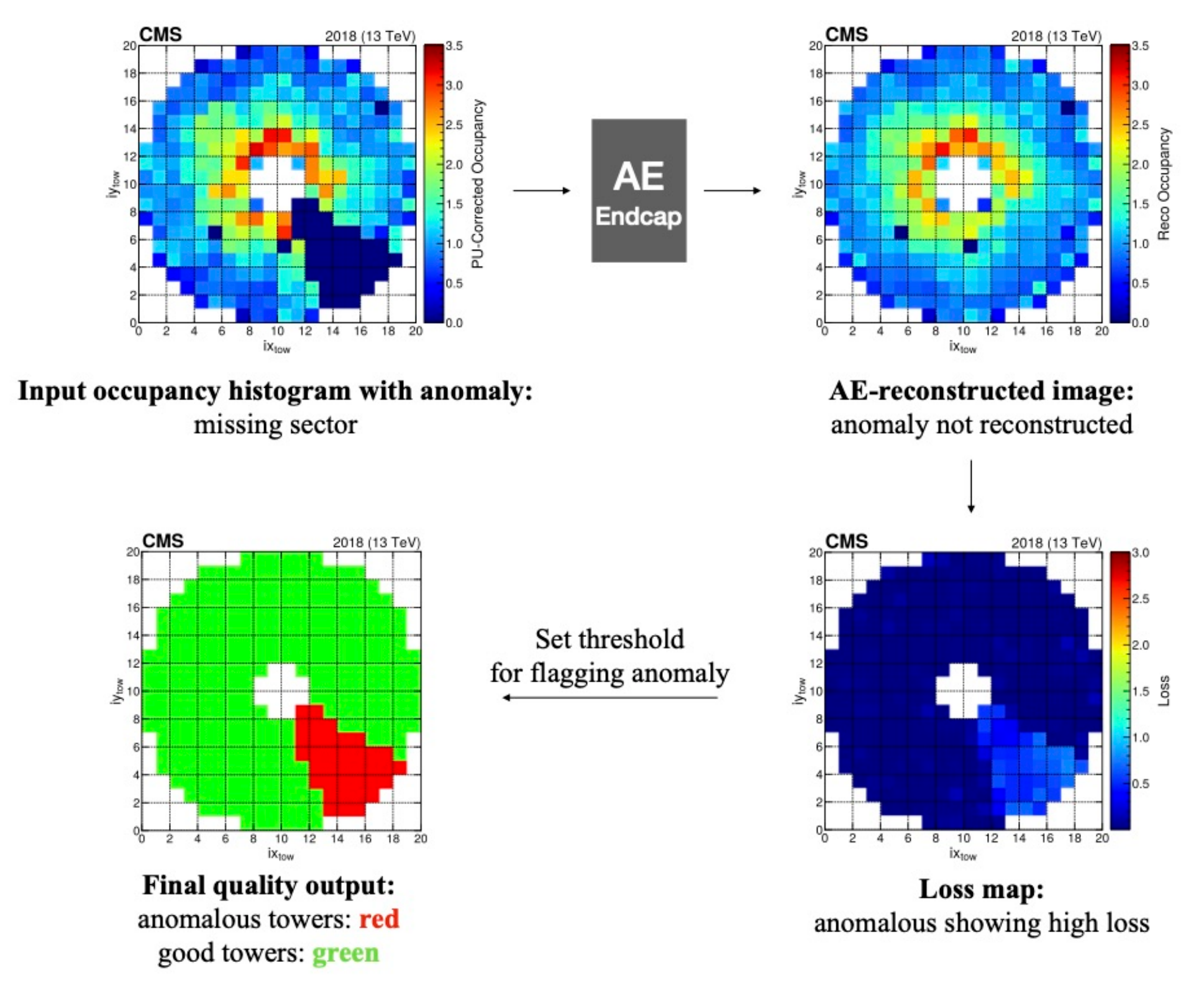}
}
    \caption{Illustration of AE-based anomaly detection strategy.}
    \label{fig:AE_strategy}
\end{figure}

A network trained on good images will learn to reconstruct them well by
minimizing this loss function. When fed with anomalous data, the AE
returns higher loss in the anomalous region, forming the basis of the
anomaly detection strategy as illustrated in
Fig.~\ref{fig:AE_strategy} using endcap images as an example.  The
input occupancy image (top left) is fed to the AE, which outputs a
reconstructed image (top right). Then the Mean Squared Error on each
tower is calculated and plotted as a 2D loss map in the same coordinates.
As shown in the bottom-right panel, the anomalous region
is highlighted with the loss higher than the rest of the image. After
applying some post-processing steps explained in
Sec.~\ref{sec:spatialcorr}, a threshold to flag the anomaly is
calculated based on the anomalous loss values. The threshold is applied
to the post-processed loss map to create a quality plot (bottom left),
where towers with the loss above the threshold are tagged as anomalous
(shown in red), while towers with loss below threshold are
identified as good (green).

\subsection{ECAL Spatial Response and Time Corrections}
\label{sec:spatialcorr}

Corrections that take into account the spatial variations in the ECAL
response and the time-dependent nature of anomalies in the detector are
implemented in order to effectively maximize the anomaly detection
efficiency while minimizing the false positive detection probability.  

\begin{figure}[tb]
\centering{
\includegraphics[width=0.23\textwidth]{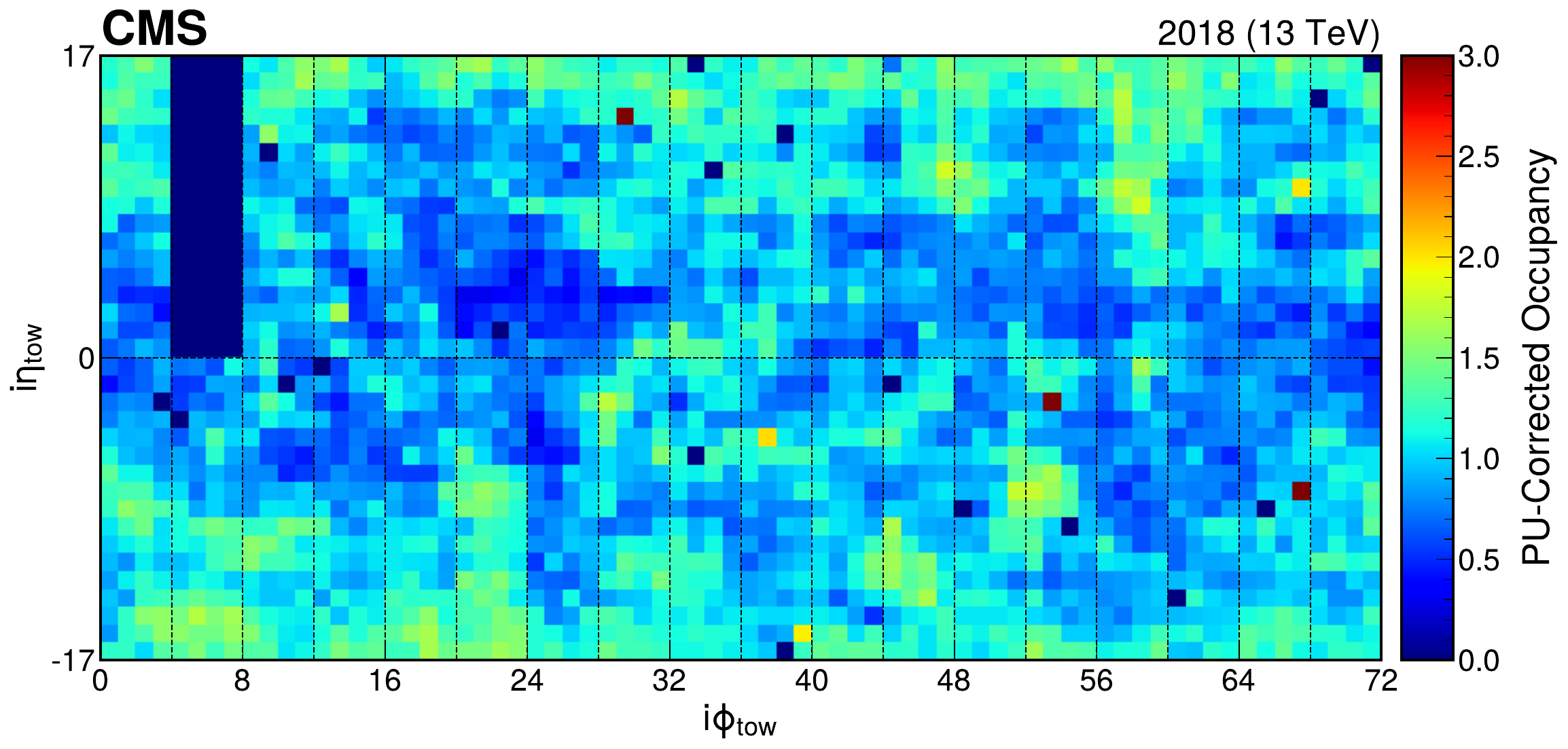} 
\includegraphics[width=0.23\textwidth]{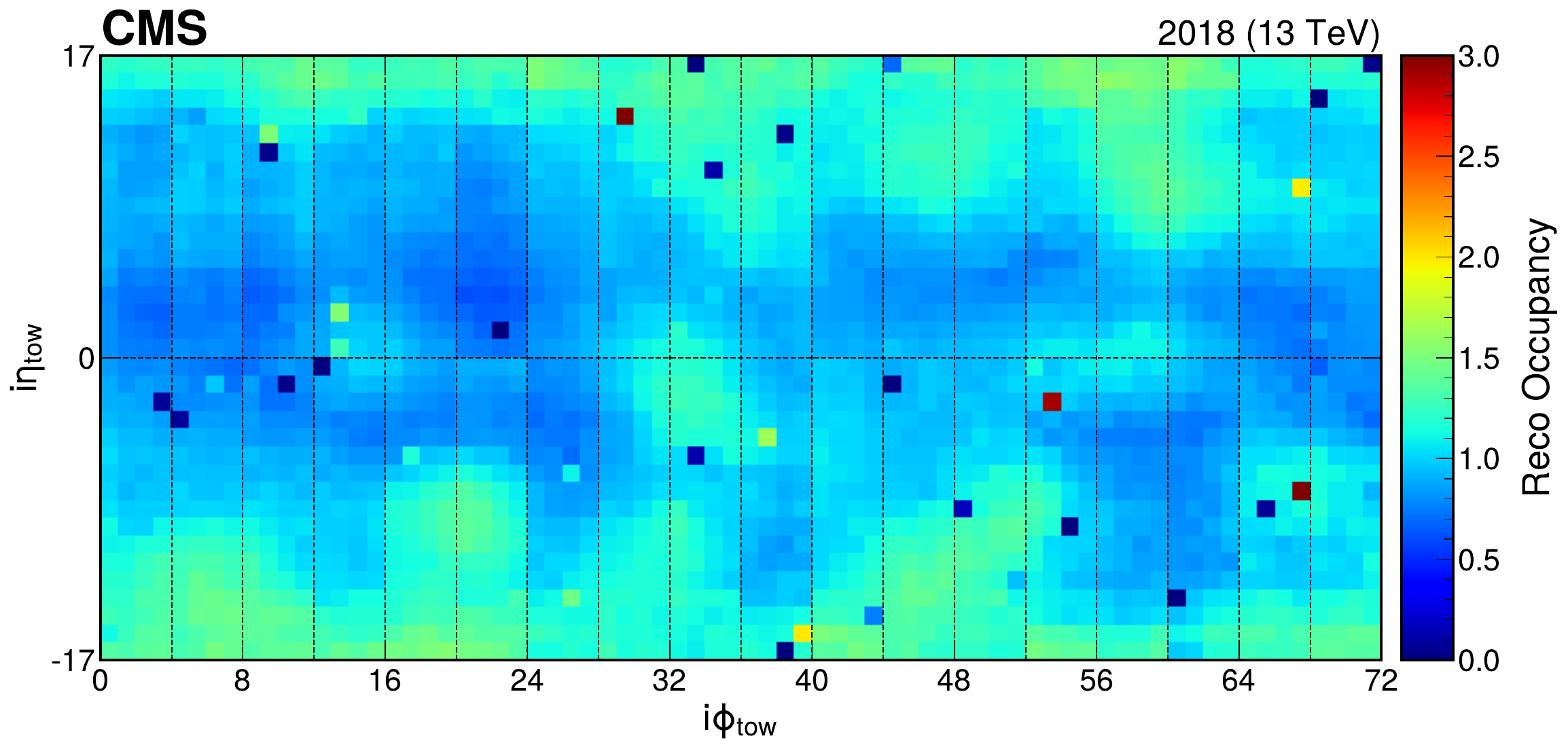} 
\includegraphics[width=0.23\textwidth]{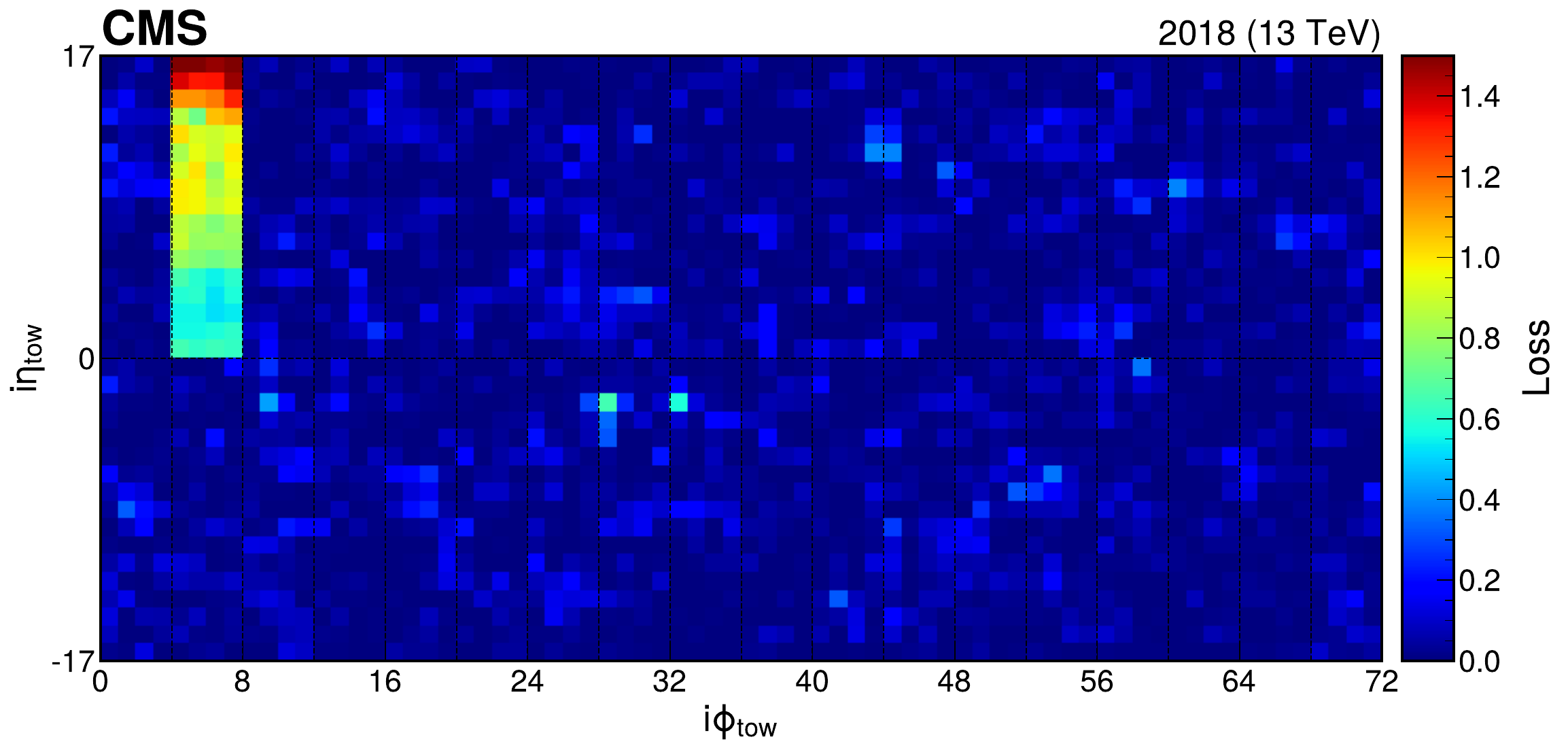} 
\includegraphics[width=0.23\textwidth]{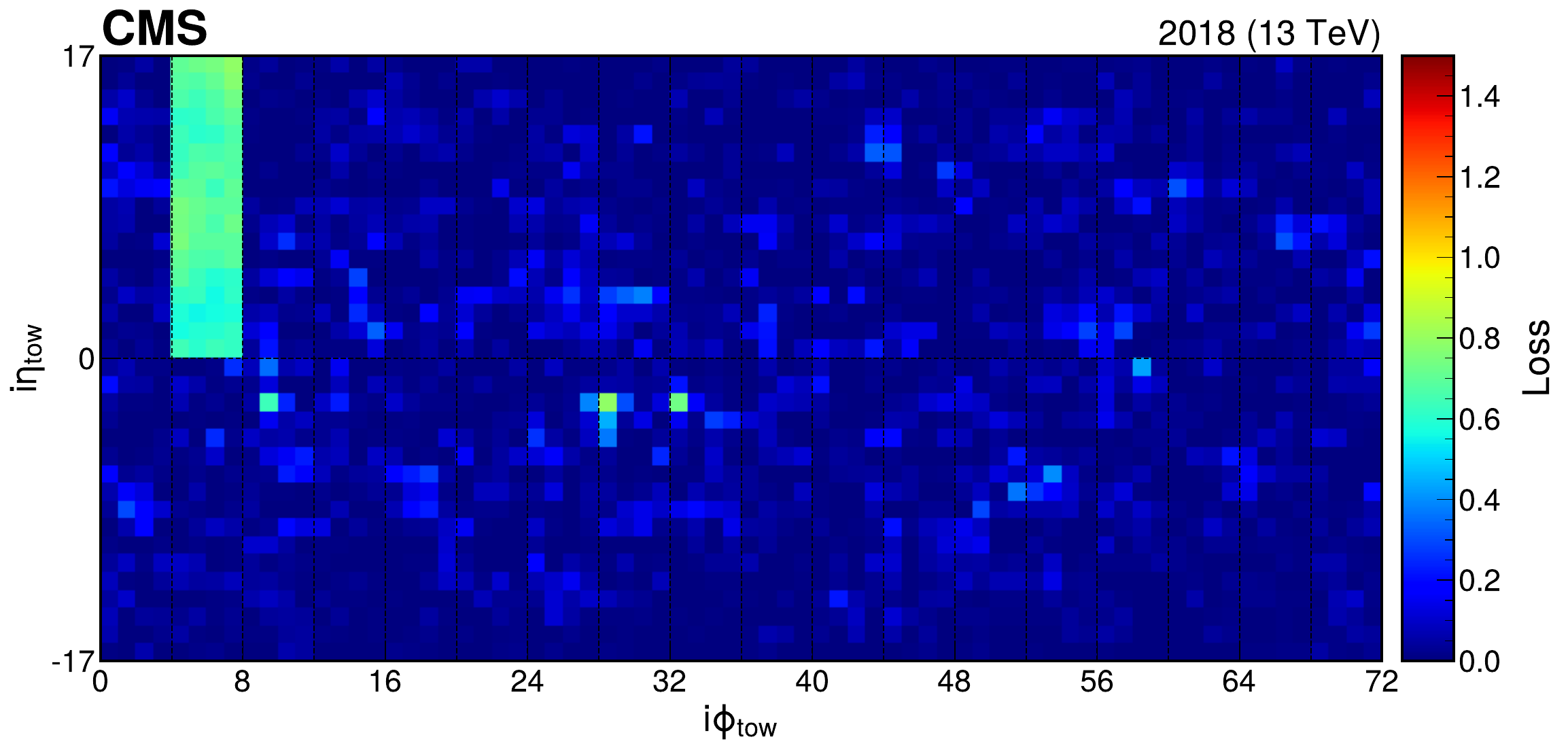}
}
\caption{Top-left: Occupancy map with a missing supermodule in the
  barrel. Top-right: AE-reconstructed occupancy map. Bottom-left: Loss map showing the
  missing supermodule, indicating higher loss at high $|\eta|$ owing to
  differences in the detector response. Bottom-right: Loss map after the spatial
  correction. } 
\label{fig:SM}
\end{figure}

Since the multiplicity of particle production in a fixed rapidity
interval is constant at a hadron collider, the number of particles per
geometric interval increases for higher~$|\eta|$, which is related to
rapidity. Thus, it is observed that ECAL crystals in regions of
high~$|\eta|$ exhibit higher occupancy than those of low~$|\eta|$ in
both the barrel and endcaps.  This difference in detector response is
also visible in the AE loss map for specific anomalies as illustrated in
Fig.~\ref{fig:SM} with a missing supermodule consisting of 68 sets of
5$\times$5 crystals. The top left plot shows the occupancy map with one
supermodule having zero occupancy. The figure to the right reflects the
corresponding AE-reconstructed output where the AE fails to reconstruct
the anomaly. The bottom left diagram of Fig.~\ref{fig:SM} is the
tower-level loss map calculated between the input and output, exhibiting
high loss in the anomalous supermodule region; the towers at the highest
$|\eta|$ tend to have a higher loss than those at lower $|\eta|$ due to
the higher average occupancy in these regions. To mitigate this effect
and obtain uniform loss in the anomalous region, the loss is normalized
by the average occupancy indicated in the top-left of
Fig.~\ref{fig:SM}. After this ``spatial response correction'', flat loss
is observed in the anomalous region as seen in the bottom right of
Fig.~\ref{fig:SM}.

Real anomalies persist with time in consecutive LSs, while random
fluctuations average out. An additional correction is implemented to
exploit the time-dependent nature of real anomalies,
named ``time correction'', which brings a significant improvement in the
AE performance.  Spatially corrected loss maps from three consecutive
LSs are multiplied together at the tower level.  The resulting
time-multiplied loss map typically shows that the persistent anomaly of
a real anomaly such as a missing supermodule is enhanced and random
fluctuations from each LS are suppressed reducing false positives.
It is observed that multiplication rather than
averaging is a better strategy for enhancing and suppressing the
resulting loss values.
%

\section{Results}
\label{sec:Results}

\subsection{Anomaly Tagging Threshold and Performance Metric}
\label{sec:FDR}

\begin{figure}[tb]
\centering{
\includegraphics[width=0.45\textwidth]{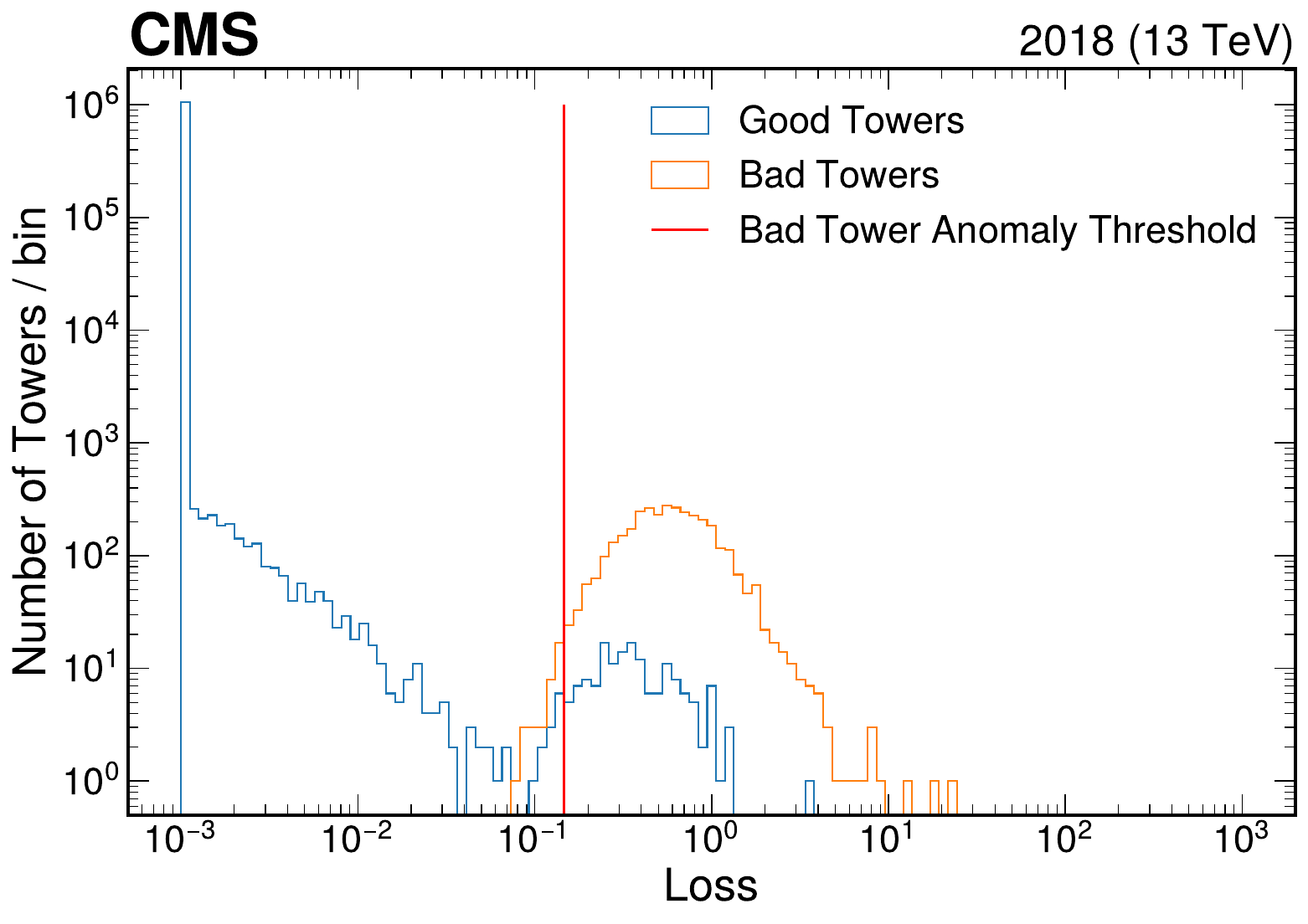} }
\caption{Loss distribution for zero occupancy tower scenario after spatial and time correction for EE$-$. The anomaly threshold is set as the lower 1\% of the zero occupancy tower loss values. }
\label{fig:loss_thres}
\end{figure}

The goal of the ML-based DQM system is to maximize the anomaly detection
efficiency while minimizing the number of false positives.  An anomaly
is tagged using a threshold obtained from a validation set with fake
anomalies. The threshold on the final post-processed loss map is chosen
such that the loss values of 99\% of anomalous towers are above the
threshold as illustrated in Fig.~\ref{fig:loss_thres} showing the
loss distribution from a zero occupancy tower scenario.

To assess the performance of the AE network, the False Discovery
Rate~(FDR) is used as a metric: 
\begin{equation}
    \textrm{FDR} = \frac{\textrm{no.~good towers above anomaly threshold}}{\textrm{no.~good $+$ bad towers above threshold}}
\label{eq:FDR1}
\end{equation}
The FDR value for 99\% anomaly detection represents the fraction of
false detection in all anomalies detected, when using the threshold
chosen to catch 99\% of the anomalies present in the dataset. In other
words, the FDR is the ratio of good towers tagged as anomalous to all
towers labeled as anomalous by the AE.  A lower FDR indicates better
performance and fewer false alarms during data taking.

\subsection{Testing on Fake Anomalies}
\label{sec:fake}

The performance of the AE-based DQM method is studied first on three
distinct anomaly scenarios -- missing supermodule/sector, single zero
occupancy tower, and single hot tower -- where artificial (fake)
anomalies are added onto good images.  Tables~\ref{tab:barrel} and
\ref{tab:endcap} summarize the FDR values calculated with anomaly
tagging thresholds determined for each scenario for 99\% anomaly
detection. For both the barrel and the endcaps, the FDRs for the single
zero occupancy tower scenario are observed to be always higher than those
for the single hot tower case.  This is because hot towers are in
general easier to spot, as they stand out with much higher occupancy
compared to neighboring towers of average occupancy.

\begin{table}[tb]
\centering{
\caption{Summary of FDR using 99\% anomaly detection threshold for the ECAL barrel fake anomaly scenarios.}
\resizebox{0.48\textwidth}{!}{\begin{tabular}{|c|c|c|c|}
\hline
\multicolumn{1}{|l|}{} & \multicolumn{3}{c|}{FDR for 99\% anomaly detection}   \\
\hline
& Missing & Zero Occup. & Hot \\
& Supermodule & Tower & Tower \\
                                \hline
\begin{tabular}[c]{@{}c@{}}AE \\ no correction \end{tabular}      & 3.6\%      & 51\%                            & 2.8\%                          \\ \hline
\begin{tabular}[c]{@{}c@{}}AE after\\  spatial correction\end{tabular}      & 3.1\%      & 49\%                            & 2.9\%                          \\ \hline
\begin{tabular}[c]{@{}c@{}}AE after\\  spatial and \\ time corrections\end{tabular}       & 0.13\%     & 4.1\%                           & $<$ 0.01\%                          \\ \hline
\end{tabular}}
\label{tab:barrel}
}
\end{table}

\begin{table}[tb]
\centering{
\caption{Summary of FDR using 99\% anomaly detection threshold for fake anomaly scenarios in the endcaps.}
\resizebox{0.48\textwidth}{!}{\begin{tabular}{|c|c|c|c|c|c|c|}
\hline
\multicolumn{1}{|l|}{} & \multicolumn{6}{c|}{FDR for 99\% anomaly detection}   \\ \hline
\multicolumn{1}{|c|}{}                                    & \multicolumn{2}{c|}{Missing Sector} & \multicolumn{2}{c|}{Zero Occup.~Tower} & \multicolumn{2}{c|}{Hot Tower} \\ \hline
 & EE$+$ & EE$-$ & EE$+$ & EE$-$ & EE$+$ & EE$-$ \\ \hline
\begin{tabular}[c]{@{}c@{}}AE \\  no correction\end{tabular}      & 29\%      & 28\%                            & 86\%      & 86\%         & $<$ 0.01\%      & $<$ 0.01\%       \\ \hline
\begin{tabular}[c]{@{}c@{}}AE after\\  spatial correction\end{tabular}      & 1.8\%     & 2.2\%       & 11\%     & 14\%                      & 0.02\%               & 0.04\% \\
\hline
\begin{tabular}[c]{@{}c@{}}AE after\\  spatial and \\ time corrections\end{tabular}      & 0.06\%      & 0.18\%                            & 1.4\%      & 4.4\%     & $<$ 0.01\%      & $<$ 0.01\%     \\ \hline
\end{tabular}}
\label{tab:endcap}
}
\end{table}

The effect of each consecutive correction on the FDRs can be seen from
the tables.  The AE spatial correction reduces the FDRs in the missing
supermodule/sector and the single zero occupancy tower scenarios, where
the occupancy values are set to zero for the barrel/endcaps.  Without
the correction, the loss values for the towers with zero occupancy
anomalies are proportional to the towers' nominal occupancy, which
indicates that the loss is biased to be larger in the higher~$|\eta|$
region (see, e.g., Fig.~\ref{fig:SM}). The spatial correction has a
greater effect for the endcaps than for the barrel, as the gradient in
occupancy values across the towers is more pronounced for the endcaps.
In the case of the hot tower anomaly, the FDRs increase after the
spatial correction. This is because the hot tower loss is biased to be
higher in the opposite direction, towards the lower~$|\eta|$ region.
However, this effect is mitigated by the time correction that greatly
improves the FDRs for all anomaly scenarios, with excellent final
performance scores for both the barrel and the endcaps.  


\subsection{Testing on Real Anomalies and Deployment}
\label{sec:realanom}

\begin{figure}[tb]
\centering{
  \includegraphics[width=0.24\textwidth]{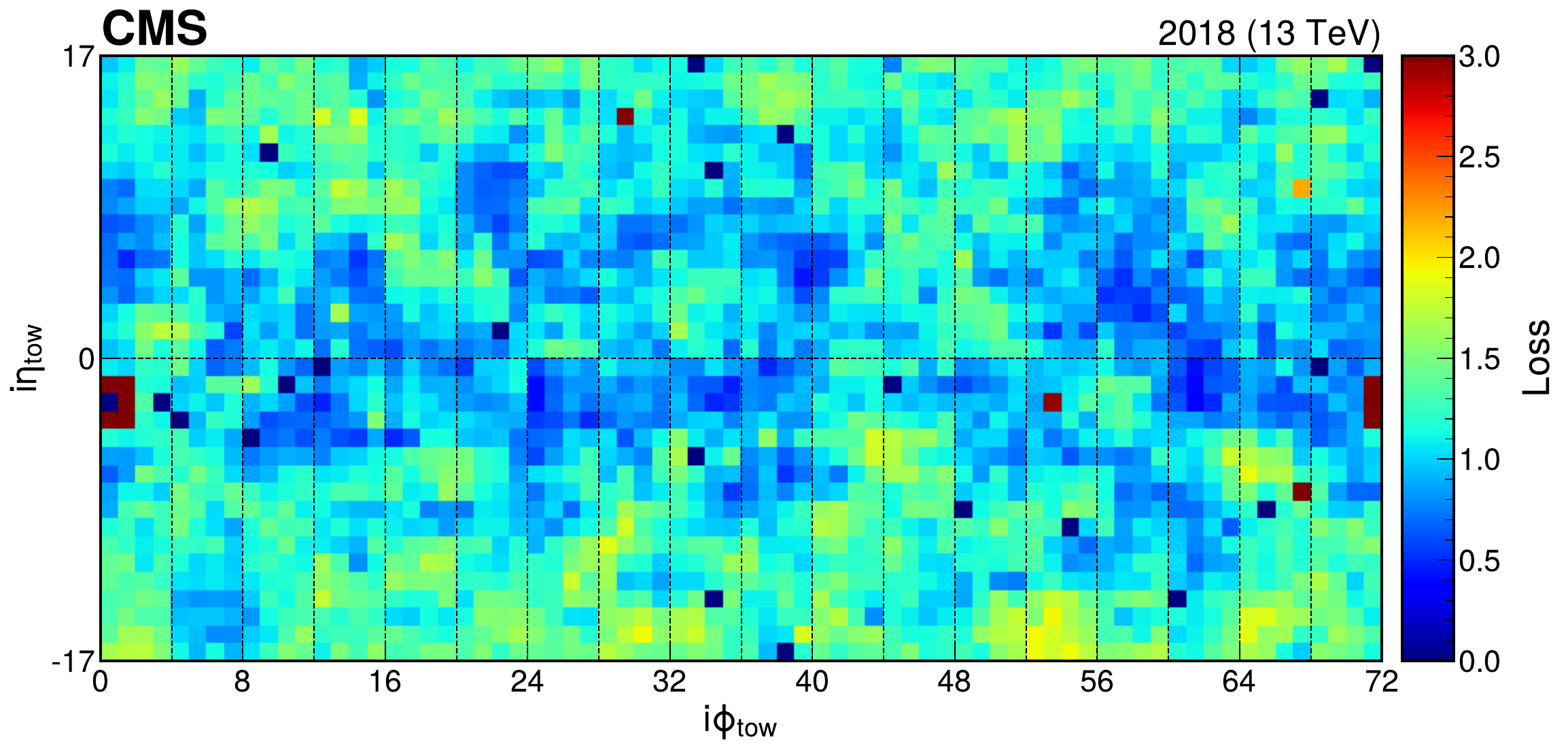}
  \includegraphics[width=0.22\textwidth]{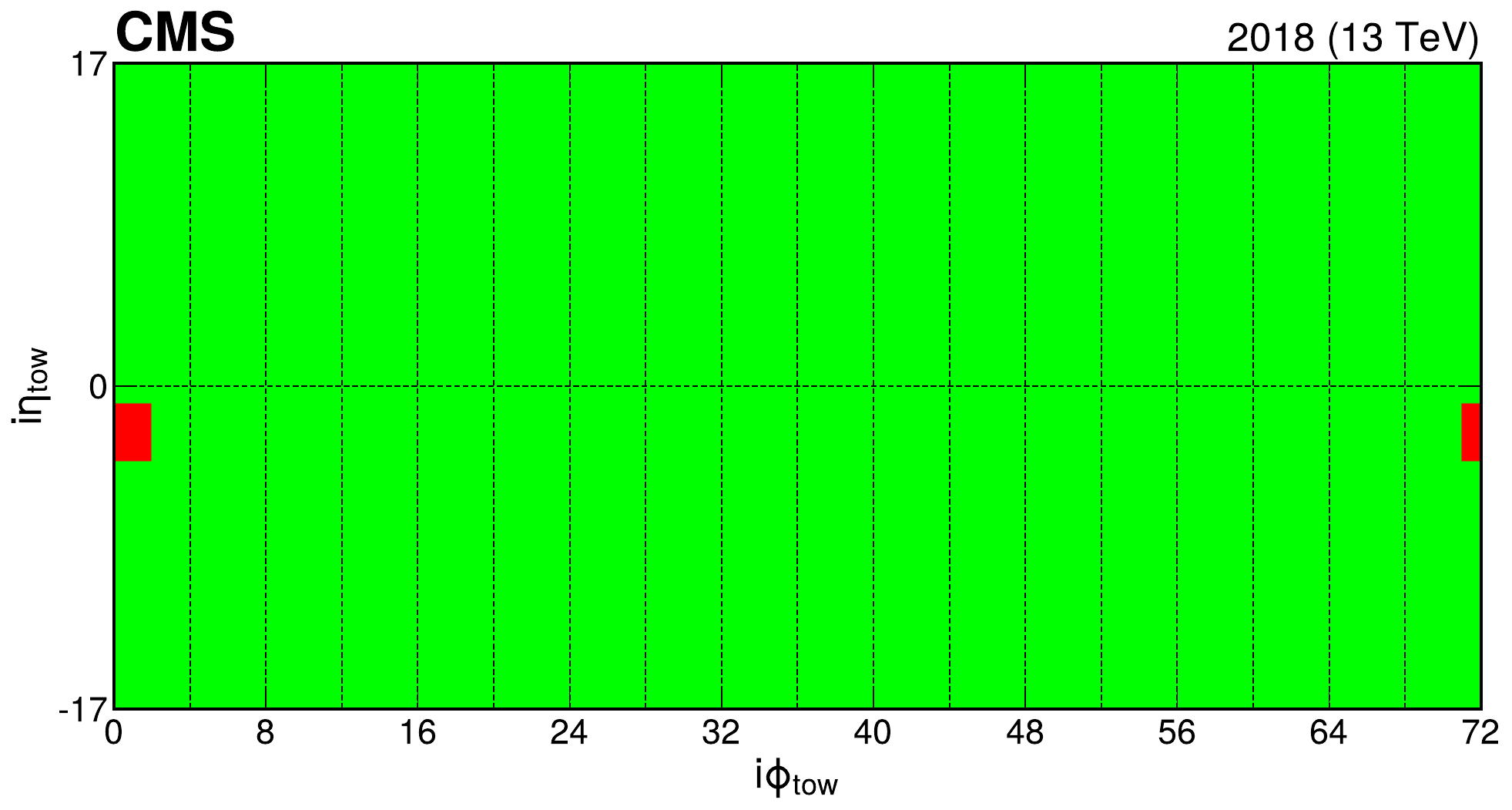}
}
\caption{
Input occupancy images with real anomalies and corresponding AE quality
plots from a 2018 run with hot towers. }
\label{fig:realanom_EB}
\end{figure}

Following the tests on fake anomalies, the AE performance is studied on
known anomalous data from LHC runs in 2018 and 2022. The input occupancy
images with anomalies and the final quality plots from the AE loss maps
are illustrated in Fig.~\ref{fig:realanom_EB} showing a barrel
occupancy map with a region of hot towers and a zero occupancy tower in
the center from a 2018 run (left). The AE quality output on the right-hand
side of Fig.~\ref{fig:realanom_EB} correctly identifies all the
anomalous towers shown in red. It is interesting to note that this error
was not detected in the online DQM global quality plots at the time of
data taking, while the AE is able to detect it.

\begin{figure}[t]
\centering{
  \includegraphics[width=0.22\textwidth]{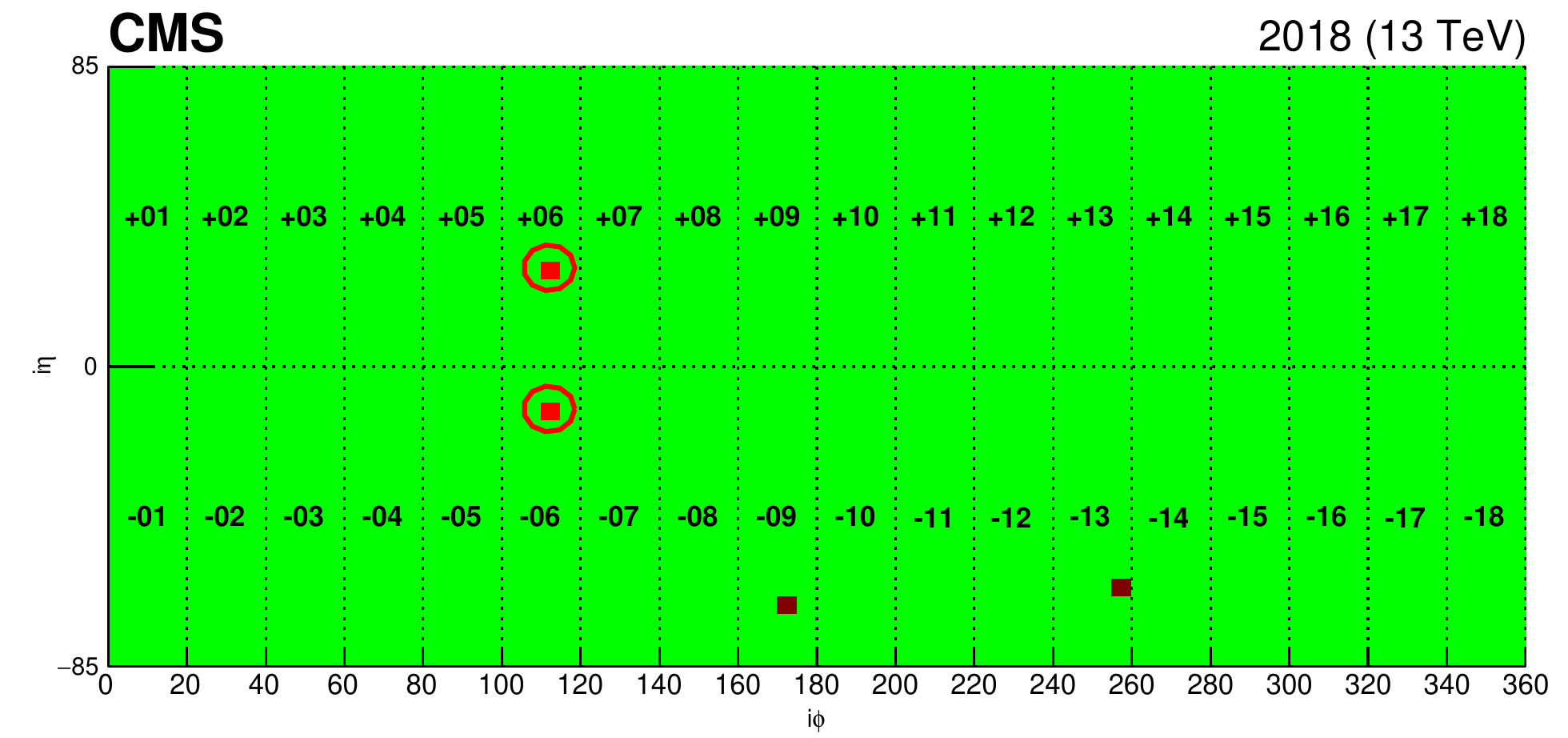}
  \includegraphics[width=0.24\textwidth]{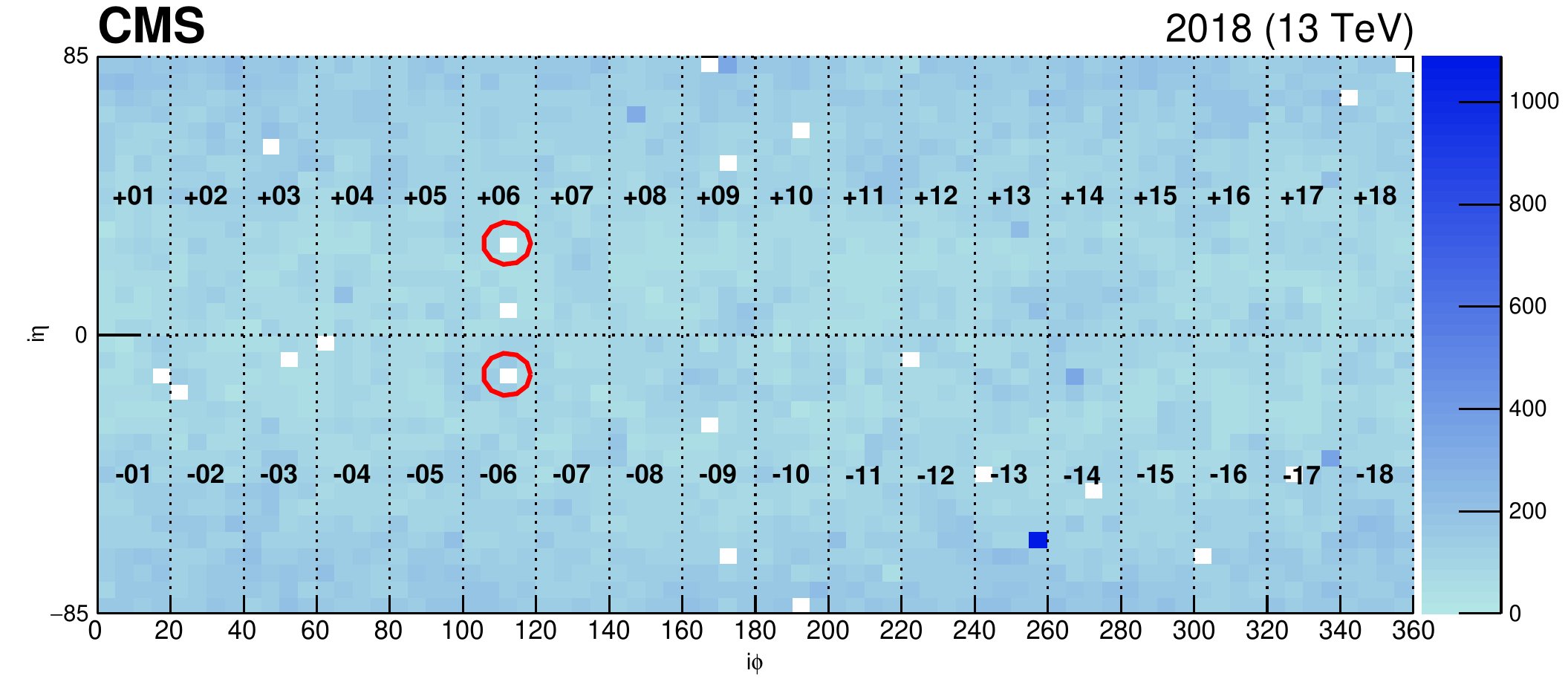}
}
\caption{
Left: From a 2022 Run ML quality plot in the ECAL DQM from the AE model,
with the new bad towers circled. Right: Occupancy plot of 1 LS. } 
\label{fig:EB-deploy1}
\end{figure}

The AE-based anomaly detection system labeled MLDQM has been deployed in
the CMS ECAL online DQM workflow for the barrel starting in LHC
Run\,3 in 2022 and for the endcaps in 2023. New ML quality plots from
the AE (see Fig.~\ref{fig:EB-deploy1}) have been added to the ECAL
DQM. The model inference is accomplished using the trained Pytorch
models exported to ONNX~\cite{onnx}, which is implemented in the CMS
software framework using ONNX Runtime.
The MLDQM models have shown so far very good performance with Run\,3
data. As an example, Fig.~\ref{fig:EB-deploy1} on the left illustrates
the new quality plot obtained from the inference of the trained AE model
for the barrel, using the occupancy histogram shown in
Fig.~\ref{fig:EB-deploy1} on the left as input to the model.  
Two circled red
towers can be seen in the supermodules~EB+06 and EB$-$06, both
corresponding to zero occupancy towers in the input occupancy map as
shown on the right of Fig.~\ref{fig:EB-deploy1}.

\section{Summary}
\label{sec:summary}

A production-level AE based anomaly detection and localization system
has been developed for the CMS ECAL using semi-supervised machine
learning. This work was just published in Ref.~\cite{CMSECAL:2023fvz}.
The anomaly detection system using machine learning described in this
paper can be generalized and adapted not only to other subsystems of the
CMS detector but also to other particle physics experiments for anomaly
detection and data quality monitoring.

\subsection*{Acknowledgements}

We congratulate our colleagues in the CERN accelerator departments for
the excellent performance of the LHC and thank the technical and
administrative staffs at CERN and at other CMS institutes for their
contributions to the success of the CMS effort.  In addition, we
gratefully acknowledge the computing centres and personnel of the
Worldwide LHC Computing Grid and other centres for delivering so
effectively the computing infrastructure essential to our analyses.
This research is supported in part by the U.S. Department of Energy,
Office of Science, through DOE award DE-SC0010118.

%
\bibliography{CALOR_AE_ECAL_DQM.bib} 

\begin{thebibliography}{8}

\bibitem{CMS:2008xjf}
S.~Chatrchyan et~al. (CMS), The {CMS} experiment at the {CERN} {LHC}, JINST
  \textbf{3}, S08004 (2008). \doiwoc{10.1088/1748-0221/3/08/S08004}

\bibitem{DQM2019_Azzolini}
V.~Azzolini et~al., {The Data Quality Monitoring software for the CMS
  experiment at the LHC: past, present and future}, EPJ Web Conf. \textbf{214},
  02003 (2019). \doiwoc{10.1051/epjconf/201921402003}

\bibitem{ML_whitepaper}
K.~Albertsson et~al., {Machine Learning in High Energy Physics Community White
  Paper}, J. Phys. Conf. Ser. \textbf{1085}, 022008 (2018),
  \texttt{1807.02876}. \doiwoc{10.1088/1742-6596/1085/2/022008}

\bibitem{Nachman-anomalydetection}
B.~Nachman, {Anomaly Detection for Physics Analysis and Less than Supervised
  Learning} (2020), \texttt{arXiv:2010.14554}.

\bibitem{AE}
G.E. Hinton, R.R. Salakhutdinov, Reducing the dimensionality of data with
  neural networks, Science \textbf{313}, 504 (2006).
  \doiwoc{10.1126/science.1127647}

\bibitem{lecun:98}
Y.~Lecun et~al., Gradient-based learning applied to document recognition,
  Proceedings of the IEEE \textbf{86}, 2278 (1998). \doiwoc{10.1109/5.726791}

\bibitem{onnx}
J.~Bai, F.~Lu, K.~Zhang et~al., Onnx: Open neural network exchange,
  \url{https://github.com/onnx/onnx} (2019)

\bibitem{CMSECAL:2023fvz}
D.~Abadjiev et~al. (CMS ECAL), {Autoencoder-Based Anomaly Detection System for
  Online Data Quality Monitoring of the CMS Electromagnetic Calorimeter},
  Comput. Softw. Big Sci. \textbf{8}, 11 (2024), \texttt{arXiv:2309.10157}.
  \doiwoc{10.1007/s41781-024-00118-z}

\end{thebibliography}

\end{document}